\documentstyle[12pt]{article}
\begin{document}

\title{\bf Bifurcation Curves of Limit Cycles
in some Li\'{e}nard Systems}
\author{R. L\'{o}pez-Ruiz$^{\ddag}$ and J.L. L\'{o}pez$^{\dag}$ \\
                                  \\
{$^{\ddag}$\small Departamento de F\'{\i}sica Te\'{o}rica} \\
{\small Facultad de Ciencias, Universidad de Zaragoza,
50009-Zaragoza (Spain)} \\
{$^{\dag}$\small Departamento de Matem\'{a}tica Aplicada} \\
{\small Facultad de Ciencias, Universidad de Zaragoza,
50009-Zaragoza (Spain)}
\date{ }}

\maketitle
\baselineskip 8mm
 \begin{center} {\bf Abstract} \end{center}
Li\'{e}nard systems of the form $\ddot{x}+\epsilon f(x)\dot{x}+x=0$,
with $f(x)$ an even continous function, are considered.
The bifurcation curves of limit cycles are calculated exactly
in the weak ($\epsilon\rightarrow 0$) and in the strongly
($\epsilon\rightarrow\infty$) nonlinear regime in some examples.
The number of limit cycles does not increase
when $\epsilon$ increases from zero to infinity in all the cases analyzed.
$\;$\newline

\newpage
\section{Introduction}

Self-sustained oscillations are found very often in nature.
There are many examples in different branches of science such as
in biology, chemistry, mechanics, electronics, fluid dynamics, etc.
\cite{andronov,hao}. Nonlinearities are required in order to have this
kind of behaviour. The system reaches an oscillatory dynamics caracterized
by a preferred period, wave form and amplitude, stable under slight
perturbations. The oscillations are generated by an internal balance
of amplification and dissipation, even in the absence of external
periodic forcing. (For instance, a nonlinear damping force which
increases the amplitude for small velocities and decreases
it for large velocities). This dynamical state can be modelled
by the stable limit cycles found in specific nonlinear differential
equations.

Limit cycles are isolated closed trajectories in phase space.
They are stable if the neighbouring solutions
tend to them in an asymptotic sense or unstable if the
neighbouring solutions unwind from them.
Determination of the number, amplitude and
loci  of limit cycles in a general
nonlinear system is an unsolved problem that has attracted much
attention in this century \cite{yan}. This constitues a part of the
Hilbert's Sixteenth Problem \cite{hilbert} when we are restricted
to two-dimensional autonomous systems of the form:
\begin{eqnarray}
 \dot{x} & = & P_n(x,y), \nonumber \\
 \dot{y} & = & Q_n(x,y),
\end{eqnarray}
where $P_n$ and $Q_n$ are polynomials of degree $n$ with real
coefficients. Althoug it has been proved that the number
of limit cycles in systems (1) is finite \cite{ecalle,ilya},
the determination of the maximal number $H_n$ of limit
cycles is still far away of being known.

The van der Pol oscillator $\ddot{x}+\epsilon (x^2-1)\dot{x}+x=0$,
where $\dot{x}(t)=dx(t)/dt$, is an example of system (1) that has
been exhaustively studied. In this case, $P_3(x,y)=y$ and
$Q_3(x,y)=-\epsilon (x^2-1)y-x$. It displays a limit cycle whose
uniqueness and non-algebraicity has been shown for the whole range
of the parameter $\epsilon$. Its behaviour runs from near-harmonic
oscillations for $\epsilon$ close to zero $(\epsilon\rightarrow
0)$ to relaxation oscillations when $\epsilon$ tends to infinity
$(\epsilon\rightarrow\infty)$, making it a good model for many
practical situations \cite{vanderpol,lopez}.

A generalization of the van der Pol oscillator
is the Li\'{e}nard equation,
\begin{equation}
 \ddot{x}+\epsilon f(x)\dot{x}+x=0,
\end{equation}
with $\epsilon$ a real parameter and $f(x)$ any real function.
When $f(x)$ is a polynomial
of degree $N=2n+1$ or $2n$ this equation of the form (1)
with $P_{N+1}(x,y)=y$ and $Q_{N+1}(x,y)=-\epsilon f(x)y-x$.
It has been conjectured by Lins, Melo and Pugh (LMP-conjecture)
that the maximum number of limit cycles allowed is just $n$ \cite{lins}.
It is true if $N=2$, or $N=3$
or if $f(x)$ is even and $N=4$ \cite{lins,rychkov}.
Also, it is true in the strongly nonlinear regime
$(\epsilon\rightarrow\infty)$ when $f(x)$ is an even
polynomial \cite{lopez1}. There are no general results
about the limit cycles when $f(x)$ is a polynomial
of degree greater than $5$ neither, in general, when $f(x)$
is an arbitrary real function \cite{giacomini,giacomini1}.

In the present paper, we are interested in the Li\'{e}nard
equation when $f(x)$ is a continous even function, otherwise arbritary.
We exploit the fact that the calculation of the number of limit cycles
in the weak $(\epsilon\rightarrow 0)$ and in the strongly
$(\epsilon\rightarrow\infty)$ nonlinear regimes is possible
for this kind of functions by means of simple algorithms.
In fact, we find exactly the bifurcation
curves of limit cycles in both regimes
for several examples of viscous terms $f(x)$.
Section 2 is devoted to explain
the strategies (or algorithms) used
to calculate the amplitude and number
of limit cycles in those extreme regimes,
and in Section 3 we analyze some particular cases
found in the literature. Last Section contains the
conclusions.

\section{Limit Cycles in the Li\'{e}nard Equation}

In order to study the limit cycles of equation (2) it is convenient
to rewrite it in the coordinates $(x,\dot{x})=(x,y)$ in the plane.
We perform the change of variables $\dot{x}(t)=y(x)$ and
$\ddot{x}(t)=y(x)y'(x)$ (where $y'(x)=dy/dx$):
\begin{equation}
 yy'+\epsilon f(x)y+x = 0.
\end{equation}
A limit cycle $C_l\equiv (x,y_{\pm}(x))$ of equation (3)
has a positive branch $y_+(x)>0$ and a negative branch $y_-(x)<0$.
They cut the $x$-axis in two points $(-a_1,0)$ and $(a_2,0)$
with $a_1,a_2>0$. The oscillation $x$ runs in the interval
$-a_1<x<a_2$. \newline
The origin $(0,0)$ is the only fixed point of equation (3).
Then every limit cycle $C_l$ solution of Eq. (3) encloses
the origin. The result is a nested set of closed curves
that defines the qualitative distribution
of the integral curves in the plane $(x,y)$. The stability
of the limit cycles is alternated. For a given stable limit cycle,
the two neighbouring limit cycles, the closest one in its interior
and the closest one in its exterior, are unstable,
and viceversa (Fig. 1).

When $f(x)$ is an even function, the symmetries of the equation (3)
impose some properties over the shape of the limit cycles.
Thus the {\it inversion symmetry} $(x,y)\leftrightarrow (-x,-y)$
implies $y_+(x)=-y_-(-x)$ and $a_1=a_2=a$. Therefore,
we can restrict ourselves to the positive branches of the
limit cycles $(x,y_+(x))$ with $-a\leq x\leq a$.
The amplitude of oscillation $a$ identifies the limit cycle.
The {\it parameter inversion symmetry}
$(\epsilon,x,y)\leftrightarrow (-\epsilon,x,-y)$
implies that if $C_l\equiv (x,y_{\pm}(x))$ is a limit
cycle for a given $\epsilon$, then
$\overline{C}_l\equiv (x,-y_{\mp}(x))$ is a limit cycle
for $-\epsilon$. Moreover if $C_l$ is stable (or unstable)
then $\overline{C}_l$ is unstable
(or stable, respectively). Therefore it is enough
to consider the positive $y$-branch $y_+(x)$ of the limit
cycles when $\epsilon>0$ for obtaining all the periodic solutions.
(The limit cycles for a given
$-\epsilon <0$ are obtained from a reflection over
the $x$-axis of those limit cycles obtained for $\epsilon>0$).

Another property of a limit cycle can be derived from
the fact that the mechanical energy $E=(x^2+y^2)/2$ is
conserved in a half oscillation:
\begin{displaymath}
\int_{-a}^{a}\frac{dE}{dx} dx = 0.
\end{displaymath}
Thus, if equation (3) is integrated along the positive branch
$y_+(x)$ of a limit cycle,
between the maximal amplitudes of oscillation,
we obtain:
\begin{equation}
\int_{-a}^{a}f(x)y_+(x) dx = 0.
\end{equation}
The solutions $y_+(x)$ of equation (3)
and (4), vanishing in the extremes,
constitute the finite set of limit cycles of equation (3).

\subsection{The Weakly Nonlinear Regime}

Li\'{e}nard system (3) reduces to the simple harmonic oscillator
when $\epsilon =0$. All the circles $y(x)=\sqrt{r^2-x^2}$
of radius $r$ about the origin are solutions. This path-diagram
is destroyed when $\epsilon$ is slightly perturbed. Only
the limit cycles survive as closed curves. They will have
a slightly modified circular form. At order zero in $\epsilon$,
we can suppose them as circles $y_+(x)=\sqrt{a^2-x^2}$
with different amplitudes $a$'s. Obviously, at this order,
the condition (3) is verified, and condition (4) reads:
\begin{equation}
\beta(a)\equiv\int_{-a}^{a}f(x)\sqrt{a^2-x^2} dx = 0.
\end{equation}
Each solution $\pm a$ of the equation $\beta(a)=0$ is the amplitude
of a limit cycle of the Li\'{e}nard system in the weak nonlinear
regime. And viceversa, the amplitudes of all limit cycles
of equation (3) are solutions of equation (5) in that regime.
These results are exact for $\epsilon =0$. In conclusion,
equation (5) determines the amplitudes of the
limit cycles of Li\'{e}nard system defined by
$f(x)$ when $\epsilon\rightarrow 0$.

The stability of a limit cycle in this regime is given,
at the lowest order in $\epsilon$, by the sign of the integral:
\begin{displaymath}
\sigma\equiv -\int_{-a_0}^{a_0}\frac{\epsilon f(x)}{y_+(x)}
= -\frac{\epsilon}{a_0}\left [\frac{d\beta(a)}{da}
\right ]_{a_0},
\end{displaymath}
where $a_0>0$ is a solution of equation (5).
The limit cycle is stable for $\sigma<0$
and unstable for $\sigma>0$.

As an example, we integrate equation (5) when $f(x)$ is an
even polynomial of degree $2n$:
\begin{displaymath}
f(x)=b_0+b_2x^2+b_4x^4+\cdots +b_{2n}x^{2n},
\end{displaymath}
where $b_0,b_2,b_4,\ldots,b_{2n}$ are real coefficients.
Then, only the amplitudes $a$ that satisfy
the equation:
\begin{displaymath}
\beta(a)=\frac{\pi a^2}{2}\sum_{k=0}^{n}b_{2k}\frac{(2k)!}
{4^k(k+1)!\;k!}a^{2k} = 0
\end{displaymath}
are allowed. The solution $a=0$
corresponds to the fixed point $(0,0)$
and the factor $a^2$ can be eliminated. Thus, the possible
amplitudes $a$ are the zeros of an even polynomial
of degree $2n$. There are no more than $n$ different solutions
$a>0$ and therefore, the maximun number of limit cycles in this case
is $n$. We conclude that LMP-conjecture is true in the weak
nonlinear regime. For instance,
$f(x)=x^2-1$ in the van der Pol oscillator. Then
$\beta (a)=\pi a^2(a^2-4)/8$ and the only existing
limit cycle has the amplitude $a\simeq 2$ when $\epsilon\rightarrow 0$.
It is stable if $\epsilon>0$ and unstable if $\epsilon<0$.

\subsection{The Strongly Nonlinear Regime}

An algorithm that determines the number and amplitude of the limit
cycles of Li\'{e}nard systems in the strongly nonlinear regime
has been proposed in reference \cite{lopez1}. A first approach
to the shape of limit cycles when $\epsilon\rightarrow +\infty$
shows that the positive $y$-branch,
$y_+(x)\equiv\epsilon z_i^s(x)$, of a stable limit cycle
with amplitude $a_i^s>0$ is given by:
\begin{equation}
z_i^s(x) = \left\{\begin{array}{cl}
0 & \mbox{if}\; -a_i^s\leq x\leq s_i \\
-F(x)+F(s_i) & \mbox{if}\;\;\;\;\; s_i\leq x\leq a_i^s,
\end{array}
\right.
\end{equation}
where $F(x)=\int_0^xf(t)dt$ and $s_i<0$ is
called the {\it gluing point} of the two pieces $z(x)=0$
and $z(x)=-F(x)+ F(s_i)$.
The unstable ones,
$y_+(x)\equiv\epsilon z_i^u(x)$, with amplitude $a_i^u>0$ are given by:
\begin{equation}
z_i^u(x) = \left\{\begin{array}{cl}
-F(x)+F(u_i) & \mbox{if}\; -a_i^u\leq x\leq u_i \\
 0 & \mbox{if}\;\;\;\;\; u_i\leq x\leq a_i^u,
\end{array}
\right.
\end{equation}
where $u_i>0$ is the gluing point
of the two pieces in this case.

In the remaining of this section we give a brief  skecht
of the algorithm (for a detailed discussion
see reference \cite{lopez1}). \newline
\underline{STABLE CYCLES}:
Consider the points $s^*<0$ where $F(x)$ has a positive
local maximum and find the points $a^*$ defined
by the rule:
\begin{displaymath}
a^*=\mbox{min}\left\{
 x>s^*, F(x)=F(s^*)\right\}.
\end{displaymath}
Geometrically $a^*$ represents the $x$-coordinate
of the first crossing point between the straight $z=F(s^*)$ and
the curve $z=F(x)$ in the plane $(x,z)$. If $a^*<|s^*|$
it is not possible to build the limit cycle and we can eliminate this
$s^*$ as a possible gluing point. If $a^*>|s^*|$
the point $s^*$ is a gluing point
candidate. We rename and order all the pairs  $(s^*,a^*)$
verifying this last property as $(\bar{s}_i,\bar{a}_i^s)$
with $\bar{s}_{i+1}<\bar{s}_i<0$ and collect them into the set:
\begin{displaymath}
\bar{{\cal A}}^s\equiv\left\{
(\bar{s}_i,\bar{a}_i^s),
F(\bar{s}_i)\;\;local\;\;maximum,
F(\bar{s}_i)>0, \bar{a}_i^s>|\bar{s}_i|
\right\}.
\end{displaymath}
By construction $\bar{a}_{i+1}^s>\bar{a}_i^s$.
There are two different situations when two sucessive pairs,
$(\bar{s}_i,\bar{a}_i^s)$ and  $(\bar{s}_{i+1},\bar{a}_{i+1}^s)$,
are considered: \newline
(a) $-\bar{a}_{i+1}^s<\bar{s}_{i+1}<-\bar{a}_i^s<\bar{s}_i$.
In this case it is possible to build a two-piecewise limit cycle
with the pair $(\bar{s}_i,\bar{a}_i^s)$
as indicated in Eq. (6). This pair is picked out
and renamed again as $(s_i,a_i^s)$. \newline
(b) $-\bar{a}_{i+1}^s<-\bar{a}_i^s<\bar{s}_{i+1}<\bar{s}_i$.
Now the constuction of a limit cycle derived from the pair
$(\bar{s}_i,\bar{a}_i^s)$ is not possible. This pair is rejected.

If there is only one pair $(\bar{s}_1,\bar{a}_1^s)$, we consider it
satisfies (a).

All the existing stable limit cycles can be found
comparing the pairs $i$ and $i+1$ under rules (a)-(b) and
iterating this process. All the pairs selected by
condition (a) (and renamed as $(s_i,a_i^s)$) are collected into the set:
\begin{equation}
{\cal A}^s\equiv\{(s_i,a_i^s)\}=
\left\{(\bar{s}_i,\bar{a}_i^s)\in\bar{{\cal A}}^s,
(\bar{s}_i, \bar{a}_i^s)\;\; \mbox{verifies (a)}
\right\}.
\end{equation}
The number, $l_s = card({\cal A}^s)$, of pairs
$(s_i,a_i^s)$ is the number of
stable limit cycles of the system (3).

\underline{UNSTABLE CYCLES}:
The same process can be repeated for the unstable cycles
by considering  the points $u^*>0$, where $F(u^*)$
is a positive local maximum, and
their partners $a^*$ are defined by:
\begin{displaymath}
a^*=\mbox{max}\left\{
x<u^*,
F(x)=F(u^*)\right\}.
\end{displaymath}
The gluing point candidates $u^*$ must verify
$|a^*|>u^*$. After renaming and ordering the pairs
$(u^*,a^*)$ fulfilling this last condition
as $(\bar{u}_i,\bar{a}_i^u)$ with $\bar{u}_{i+1}>\bar{u}_i>0$,
we collect them into the set:
\begin{displaymath}
\bar{{\cal A}}^u\equiv \left\{
(\bar{u}_i,\bar{a}_i^u),
F(\bar{u}_i)\;\;local\;\;maximum,
F(\bar{u}_i)>0,
|\bar{a}_i^u|>\bar{u}_i
\right\}.
\end{displaymath}
A similar algorithm as indicated above can be applied in this case
with the following modified rules: \newline
(a') $\bar{u}_i<-\bar{a}_i^u<\bar{u}_{i+1}<-\bar{a}_{i+1}^u$.
In this case there exists an unstable two-piecewise limit cycle
resulting from the pair $(\bar{u}_i,\bar{a}_i^u)$ and given in Eq. (7).
This pair is picked out and renamed $(u_i,a_i^u)$. \newline
(b') $\bar{u}_i<\bar{u}_{i+1}<-\bar{a}_i^u<-\bar{a}_{i+1}^u$. The
pair $(\bar{u}_i,\bar{a}_i^u)$ does not produce a limit cycle
and is rejected.

If there is only one pair $(\bar{u}_1,\bar{a}_1^u)$ we consider
it satisfies (a').

We iterate the process given by rules (a')-(b').
All the pairs selected by condition (a') are collected into
the set:
\begin{equation}
{\cal A}^u\equiv\{(u_i,a_i^u)\}=
\left\{(\bar{u}_i,\bar{a}_i^u)\in\bar{{\cal A}}^u,
(\bar{u}_i, \bar{a}_i^u)\;\; \mbox{verifies (a')}
\right\}.
\end{equation}
The number, $l_u = card({\cal A}^u)$, of pairs $(u_i,a_i^u)$
is the number of unstable limit cycles of system (3).
Obviously, $l_s-1\leq l_u\leq l_s+1$.

It was claimed in \cite{lopez1}
that the total number $l$ of limit cycles of Eq. (3) in the
strongly nonlinear regime is $l=l_s+l_u$, where $l_s$ and $l_u$
are the number of stable and unstable limit cycles
respectively. The amplitudes of these limit cycles are given by
the numbers $a_i^s$ and $a_i^u$, respectively. \newline

We remark also that each pair of zeros $\pm x_i$
of $f(x)$ produces at most
one limit cycle. If $f(x)$ is an even polynomial
of degree $2n$ there will be at most  $n$ limit cycles.
Therefore,
LMP-conjecture is also true in the strongly nonlinear regime.
For instance, in the van der Pol oscillator, $F(x)=-x+x^3/3$
has an unique positive local maximum at $s=-1$. The amplitude $a$
of the only existing limit cycle when $\epsilon\rightarrow\infty$
is given by the solution of the relation $F(-1)=F(a)$, that is,
$a=2$. Its shape, $y_+(x)\equiv\epsilon z(x)$,
is (up to order $\epsilon^{-2}$) given by:
\begin{displaymath}
z(x) = \left\{\begin{array}{cl}
0 & \mbox{if}\; -2\leq x\leq-1 \\
\frac{1}{3}(-x^3+3x+2) & \mbox{if}\; -1\leq x\leq 2,
\end{array}
\right.
\end{displaymath}

\section{Bifurcation Curves in some Examples}

We apply in this section the results of the former section to
particular examples that have been studied by different
authors in the literature.

{\bf Example 1:} $\underline{\bf f(x)=x^{2n}-1}$,
with $n=1,2,3,\cdots$.
This case represents a generalization of the van der Pol
oscillator \cite{vanderpol}.
It has only a limit cycle for the whole range
of the parameter $\epsilon$. \newline
(i) $\underline{\epsilon\rightarrow 0}$:
The amplitude $a_n$, $n=1,2,3,\cdots$, of this limit cycle
in the weakly nonlinear regime
is the solution of the equation $\beta (a_n)=0$. We obtain:
\begin{displaymath}
a_n = 2\sqrt[2n]{\frac{n!(n+1)!}{(2n)!}}
\end{displaymath}
If $n=1$ then $a_1=2$ (van der Pol system) and
if $n\rightarrow\infty$ the result is $a_{\infty}$=1.
That is, $1\leq a_n\leq 2$ for all values of $n$. \newline
(ii) $\underline{\epsilon\rightarrow\infty}$:
The calculation of the amplitude $a_n$ of this limit cycle
in the strongly nonlinear regime requires to find
the positive local maxima of $F(x)=\int_0^x f(t)dt$.
In this case, the only local maximum is at
$x=-1$ with the value $F(-1)= 2n/(2n+1)$.
The amplitude $a_n$ is therefore given by:
\begin{displaymath}
a_n = F^{-1}\left(\frac{2n}{2n+1}\right)>0
\end{displaymath}
In particular, $a_1=2$ and if $n\rightarrow\infty$
the amplitude is $a_{\infty}=1$. Also, in this case,
$1\leq a_n\leq 2$ for each value of $n$.\newline
(iii) $\underline{0\ll\epsilon\ll\infty}$:
Computer simulations show that the amplitude
$a_n$ of the unique limit cycle is slighty
perturbed in this regime.

{\bf Example 2:} $\underline{\bf f(x)=(x^2-1)(x^2-k)}$,
where $k$ is a real parameter (Fig. 2).
This sytem was studied
by Lloyd in Ref. \cite{lloyd}. He showed that it has
no periodic solutions if $1/5<k<5$, and he suggested
that there exist $k_*$ and $k^*$, depending on $\epsilon$,
such that there are two periodic solutions if
$0<k<k_*$ or $k>k^*$, while there are none if
$k_*<k<k^*$. Moreover, he finds that $k^*\leq (7+\sqrt{45})/2$
for some positive $\epsilon$. Here,  the values
of $(k_*,k^*)$ in the weakly, $(k_0,k^0)$, and
in the strongly, $(k_{\infty},k^{\infty})$,
nonlinear regimes are calculated.
For intermediate values of $\epsilon$ it is found numerically
that $k_{\infty}\leq k_*\leq k_0$ and
$k^0\leq k^*\leq k^{\infty}$. \newline
(i) $\underline{\epsilon\rightarrow 0}$:
The amplitudes $a_{\pm}$, solutions of the equation
$\beta(a)=0$, of the limit cycles in this regime are the
positive values of the expression:
\begin{displaymath}
a_{\pm}=\left\{(k+1)\pm\sqrt{(k+1)^2-8k}\right\}^{\frac{1}{2}}
\end{displaymath}
If $k<0$ there is only one limit cycle of amplitude $a_+$.
If $k>0$ the sign of the discriminant $\Delta=(k+1)^2-8k$
determines the number of periodic solutions. The roots
of $\Delta$ are: $k_0=3-\sqrt{8}\simeq 0.17157$ and
$k^0=3+\sqrt{8}\simeq 5.82842$. Then, if $k_0<k<k^0$,
$\Delta$ is negative and there is
no limit cycle. If $0<k<k_0$ or $k>k^0$,
$\Delta$ is positive and the system has two
periodic solutions of amplitudes $a_{\pm}$.
In $k=0$ a limit cycle of small amplitude bifurcates
from the origin after an Andronov-Hopf bifurcation.
In $k=k_0$ and $k=k^0$ the two limit cycles
appear or disappear by a saddle-node bifurcation.\newline
(ii) $\underline{\epsilon\rightarrow\infty}$:
The limit cycles in this regime are determined by the
the positive local maxima of
$F(x)= x^5/5-(k+1)x^3/3+kx$. If $k<0$
there is only a positive local maximum at $x_+=-1$
with the value $F(-1)=(2-10k)/15$. The amplitude
$a_+$ of this limit cycle is:
\begin{displaymath}
a_+=F^{-1}\left(\frac{2-10k}{15}\right)>0
\end{displaymath}
If $0<k< 1/5$ there are two positive local maxima:
one is at $x_+=-1$ and the other one is at $x_-=\sqrt{k}$.
The condition for the existence of two
limit cycles is: $F(-1)>F(\sqrt{k})$. This is verified when
$0<k<k_{\infty}=(3-\sqrt{5})^2/4\simeq 0.14589$.
The amplitudes $a_{\pm}$ of these limit cycles are:
\begin{eqnarray*}
a_+ & = & F^{-1}\left(\frac{2-10k}{15}\right)>0 \\
a_- & = & \left|max\left\{
F^{-1}\left(\frac{(10-2k)k^{\frac{3}{2}}}{15}\right)<0
\right\}\right|
\end{eqnarray*}
If $1/5<k<1$ there is only a positive local maximum
at $x=\sqrt{k}$ for which $F^{-1}(x)$ has not an antiimage.
Then there is not periodic solutions in that interval of $k$.
The same behaviour is found when $1<k<5$, but now the positive
local maximum is located at $x=1$.\newline
If $k>5$ there are two positive local maxima localized
at $x_+=-\sqrt{k}$ and at $x_-=1$. The condition for having
periodic solutions reads: $F(-\sqrt{k})>F(1)$. This condition holds
when $k>k^{\infty}=(3+\sqrt{5})^2/4\simeq 6.85410$.
The amplitudes $a_{\pm}$ of these limit cycles are:
\begin{eqnarray*}
a_+ & = & F^{-1}\left(\frac{(2k-10)k^{\frac{3}{2}}}{15}\right)>0 \\
a_- & = & \left|max\left\{F^{-1}\left(\frac{10k-2}{15}\right)<0
\right\}\right|
\end{eqnarray*}
If $k_{\infty}<k<k^{\infty}$ the system has no
periodic solutions. \newline
(iii) $\underline{0\ll\epsilon\ll\infty}$:
Numerical computations of this system suggest that the curves
$k_*(\epsilon)$ and $k^*(\epsilon)$ behave as it is shown
in Fig. 2. Thus we find that $k_{\infty}\leq k_*\leq k_0$ and
$k^0\leq k^*\leq k^{\infty}$ for every real $\epsilon$.\newline
In summary, if $k_0<k<k^0$ there is no periodic solution,
and, if $0<k<k_{\infty}$ or $k>k^{\infty}$ the system
has two limit cycles. Observe that
$k^{\infty}<(7+\sqrt{45})/2$, and the requirement
$k>(7+\sqrt{45})/2$ is not necessary
for having two limit cycles
for some $\epsilon$, as it was proposed in \cite{lloyd}.\newline
Some amplitudes $a_{\pm}$ are given in the following table:
\begin{center}
\begin{tabular}{||c||c|c||c||}
\hline
$k$ & $\epsilon=0$ & $\epsilon=1$ & $a$ \\
\hline
\hline
0 & $\sqrt{2}$ & 1.41431 & $a_+$ \\
0 & 0 & 0 & $a_-$ \\
\hline
0.16 & 1.19001 & 1.19000 & $a_+$ \\
0.16 & 0.95072 & 0.95172 & $a_-$ \\
\hline
5 & \O & \O & $a_+$ \\
5 & \O & \O & $a_-$ \\
\hline
7 & 3.29065 & 3.29504 & $a_+$ \\
7 & 2.27410 & 2.44764 & $a_-$ \\
\hline
\end{tabular}
\end{center}

{\bf Example 3:} $\underline{\bf f(x)=5x^4-3\mu x^2+\delta}$,
where $\mu$ and $\delta$ are two real parameters (Fig. 3).
This system is a generalization of Example 2. Several authors
have studied the case $\delta=1$: in Ref. \cite{rychkov},
Rychkov shows that this equation have two cycles when
$\epsilon>0$ and $\mu>2.5$; Alsholm has improved this result
lowering the bound to $\mu>2.3178$
($\mu\ge 2.3178\;\delta^{\frac{1}{2}}$), and in \cite{odani},
Odani obtained a sharper result $\mu>\sqrt{5}$.
In \cite{giacomini}, Giacomini \& Neukirch
obtain a sequence of algebraic approximations,
in the parameter plane $(\delta,\mu)$ for $\epsilon=1$,
to the bifurcation set $B_{\epsilon=1}(\delta,\mu)=0$,
where the system undergoes a saddle-node bifurcation.
Here, the bifurcation curves $B_0(\delta,\mu)=0$
and $B_{\infty}(\delta,\mu)=0$ in the weakly and in the
strongly nonlinear regimes are calculated, respectively.
We obtain $B_0(\delta,\mu)=9\mu^2-40\delta$ and
$B_{\infty}(\delta,\mu)=\mu^2-5\delta$. For intermediate
values of $\epsilon$, numerical simulations
show that the curves $B_{\epsilon}(\delta,\mu)=0$
are localized between $B_0(\delta,\mu)=0$
and $B_{\infty}(\delta,\mu)=0$ in such a way
that if $\epsilon_2>\epsilon_1$ then
$B_{\epsilon_2}(\delta,\mu)=0$ is between
$B_{\epsilon_1}(\delta,\mu)=0$ and
$B_{\infty}(\delta,\mu)=0$.
These results are in agreement with the earlier works
cited above and let us a better understanding
of the behaviour of this system
for all the values of $\delta$ and $\mu$.\newline
(i) $\underline{\epsilon\rightarrow 0}$: The amplitudes $a_{\pm}$
of the limit cycles in this regime are the positive
solutions of the equation $\beta(a)=0$, that is:
\begin{displaymath}
a_{\pm}=\left\{\frac{1}{5}\left(3\mu\pm\sqrt{9\mu^2-40\delta}
\right)\right\}^\frac{1}{2}
\end{displaymath}
If $\delta<0$ there is only one limit cycle
of amplitude $a_+$. If $\delta>0$ and $\mu<0$ there are no real
solutions for $a_{\pm}$. If $\delta>0$
and $\mu>0$, the sign of the discriminant
$B_0(\delta,\mu)=9\mu^2-40\delta$ determines  the number of
periodic solutions. If $3\mu<\sqrt{40}\;\delta^{\frac{1}{2}}$
there is none and
if $3\mu>\sqrt{40}\;\delta^{\frac{1}{2}}$ the system has two
limit cycles of amplitudes $a_{\pm}$. At $\delta=0$ a limit cycle
of small amplitude bifurcates from the origin after
an Andronov-Hopf bifurcation. For the values $(\delta,\mu)$
where $B_0(\delta,\mu)=0$ the two limit cycles appear or disappear
by a saddle-node bifurcation. \newline
(ii) $\underline{\epsilon\rightarrow\infty}$: The positive
local maxima of $F(x)= x^5-\mu x^3+\delta x$ must be found.
If $\delta<0$ there is only a positive local maximum
at $x_0=-(\frac{3\mu+\Delta}{10})^{\frac{1}{2}}/\sqrt{10}$
where $\Delta=\sqrt{9\mu^2-20\delta}$. We have
$F(x_0)=-(3\mu^2+\mu\Delta-20\delta)x_0/25$.
The amplitude $a_+$ of this limit cycle is:
\begin{displaymath}
a_+=F^{-1}\left[\left(\frac{3\mu^2+\mu\Delta-20\delta}{25}\right)
\left(\frac{3\mu+\Delta}{10}\right)^{\frac{1}{2}} \right]>0
\end{displaymath}
If $\delta>0$ and $3\mu<\sqrt{20}\;\delta^{\frac{1}{2}}$, $F(x)$
has no local maxima and the system has no periodic solutions.
If $\delta>0$ and $3\mu\geq\sqrt{20}\;\delta^\frac{1}{2}$
there are two positive local maxima localized at
$x_+=-(3\mu+\Delta)^{\frac{1}{2}}/\sqrt{10}$
and $x_-=(3\mu-\Delta)^{\frac{1}{2}}/\sqrt{10}$.
The condition for having two limit cycles is:
$F(x_+)> F(x_-)$. This follows if
$\mu>\sqrt{5}\delta^{\frac{1}{2}}$.
The amplitudes $a_{\pm}$ of the limit cycles are:
\begin{eqnarray*}
a_+ &  = & F^{-1}\left[\left(\frac{3\mu^2+\mu\Delta-20\delta}{25}\right)
\left(\frac{3\mu+\Delta}{10}\right)^{\frac{1}{2}}\right]>0 \\
a_- &  = & \left|max\left\{
F^{-1}\left[\left(\frac{-3\mu^2+\mu\Delta+20\delta}{25}\right)
\left(\frac{3\mu-\Delta}{10}\right)^{\frac{1}{2}}\right]<0
\right\}\right|
\end{eqnarray*}
If $\mu<\sqrt{5}\delta^\frac{1}{2}$ there are not periodic solutions.
Then the bifurcation curve where the system undergoes the saddle-node
bifurcation is defined by $B_{\infty}(\delta,\mu)=\mu^2-5\delta$.\newline
(iii) $\underline{0\ll\epsilon\ll\infty}$: Numerical simulations
of this system suggest that the bifurcation curves
$B_{\epsilon}(\delta,\mu)=0$ are located between $B_0(\delta,\mu)=0$
and $B_{\infty}(\delta,\mu)=0$, in such a way that if
$\epsilon_2>\epsilon_1$ then $B_{\epsilon_2}(\delta,\mu)=0$
is located between $B_{\epsilon_1}(\delta,\mu)=0$ and
$B_{\infty}(\delta,\mu)=0$. If $\mu^*(\epsilon)$ is
the solution of $B_{\epsilon}(\delta_0,\mu)=0$ for a
fixed $\delta_0$, then
$\sqrt{40\delta_0}/3\leq\mu^*(\epsilon)
\leq\sqrt{5\delta_0}$ and
$\mu^*(\epsilon_2)>\mu^*(\epsilon_1)$ if $\epsilon_2>\epsilon_1$.\newline
As an example we give some values of the amplitudes
$a_{\pm}$ of the periodic solutions for $\mu=1$:
\begin{center}
\begin{tabular}{||c||c|c||c||}
\hline
$\delta$ & $\epsilon=0$ & $\epsilon=1$ & $a$ \\
\hline
\hline
-1 & $\sqrt{2}$ & 1.40990 & $a_+$ \\
\hline
0.1 & 1.02334 & 1.02344 & $a_+$ \\
0.1 & 0.39087 & 0.39090 & $a_-$ \\
\hline
0.3 & \O & \O & $a_+$ \\
0.3 & \O & \O & $a_-$ \\
\hline
\end{tabular}
\end{center}

{\bf Example 4:} $\underline{\bf f(x)=7x^6-
5(29+b^2)x^4+3(100+29b^2)x^2-100b^2}$,
where $b$ is a real parameter (Fig. 4).
Giacomini \& Neukirch have investigated this system
in Ref. \cite{giacomini}. They find that the solutions
of the equation $F(x,b)=0$ do not give the right
qualitative amplitude-bifurcation diagram, where
$F(x,b)=x(x^2-b^2)(x^2-2^2)(x^2-5^2)$. In fact,
the plot of the roots of $F(x,b)=0$ announce the
presence of a transtricital bifurcation near
$b=2$ and $b=5$, and an Andronov-Hopf bifurcation at $b=0$.
Their method of algebraic approximations to the
bifurcation curves shows that the
supposed transcritical bifurcations are indeed
saddle-node bifurcations. They conclude that the system
can have one or three limit cycles. We confirm these results
and stablish the correct amplitude-bifurcation curves
$a_i^{\epsilon}(b)$, $i=1,2,3$, for the three limit cycles,
in the weakly, $a_i^0(b)$, and in the strongly,
$a_i^{\infty}(b)$, nonlinear regimes. The values of $b$
for which the saddle-node bifurcations occur are
calculated in those regimes. Numerical simulations show that
for intermediate values of $\epsilon$
the amplitude-bifurcation $a_i^{\epsilon}(b)$ curves and the
values of $b$ for which the saddle-node bifurcations occur
are localized in the regions bounded by the curves $a_i^0(b)$ and
$a_i^{\infty}(b)$, $i=1,2,3$. As in the former examples,
the variation of parameter $\epsilon$ does not introduce
new qualitative information in the system, and only produces slight
perturbations in the amplitude-bifurcation diagrams.\newline
(i) $\underline{\epsilon\rightarrow 0}$: Mapple
calculations allow us to solve the equation $\beta(a)=0$
for the amplitudes $a_i^0(b)$, $i=1,2,3$. The number
of positive real solutions of that cubic equation in $a^2$
is determined by the the sign of the
polynomial $\Delta (b)=-0.01784\;b^8+1.15301\;b^6-21.65794\;b^4+
132.559\;b^2-189.45$.
If $\Delta<0$ there are three limit cycles and if $\Delta>0$
there is only one. If $\Delta=0$
a saddle-node bifurcation arises in the system.
The positive roots of $\Delta(b)$ are:
$b_1^0=1.42636$, $b_2^0=2.84148$,
$b_3^0=4.17545$ and $b_4^0=6.08945$.
If $0<b<b_1^0$, $b_2^0<b<b_3^0$
or $b>b_4^0$ then $\Delta(b)<0$ and
there are three periodic solutions .
If $b_1^0<b<b_2^0$ or $b_3^0<b<b_4^0$ then $\Delta(b)>0$
and there is only one.
The amplitudes of
the limit cycles for a given $\bar{b}$ are the cuts
of the line $b=\bar{b}$ with the curves $a_1^0(b)$,
$a_2^0(b)$ and $a_3^0(b)$ (see Fig. 4).\newline
(ii) $\underline{\epsilon\rightarrow \infty}$:
Numerical calculation of show that the values of $b$
for which the system undergoes a saddle-node bifurcation
are: $b_1^{\infty}=1.21$, $b_2^{\infty}=3.49$,
$b_3^{\infty}=3.95$ and $b_4^{\infty}=6.40$.
If $0<b<b_1^{\infty}$, $b_2^{\infty}<b<b_3^{\infty}$
or $b>b_4^{\infty}$ the system has three limit cycles,
and if $b_1^{\infty}<b<b_2^{\infty}$ or
$b_3^{\infty}<b<b_4^{\infty}$ there is only one
periodic solution (see Fig. 4). \newline
(iii) $\underline{0\ll\epsilon\ll\infty}$:
Numerical computations show that the amplitude-curves
$a_i^{\epsilon}(b)$, $i=1,2,3$, are
localized in the narrow shaded region bounded by
$a_i^0(b)$ and $a_i^{\infty}(b)$ (Fig. 4a).
Remark that the behaviour of the system suggests,
once more, that the number of
limit cycles do not increase when $\epsilon$
increases (Fig. 4b).

\section{Conclusions}

Limit cycles are isolated periodic solutions of specific
nonlinear differential equations and can model self-sustained
oscillations in nature. There are two difficult and connected
problems in relation with limit cycles: the determination
of bifurcation curves of these solutions in the parameter space
and the determination of the maximal number of such solutions.
In this work, we have exploited the possibility
of calculating the bifurcation
curves of the limit cycles of Li\'{e}nard equation
$\ddot{x}+\epsilon f(x)\dot{x}+x=0$ in the
weakly ($\epsilon\rightarrow 0$) and in the strongly
($\epsilon\rightarrow\infty$) nonlinear regimes
when the viscous term $f(x)$ is even. Firstly,
these calculations allow us to improve the results existing
in the literature for different examples in these regimes.
Secondly,
the systems analyzed seem to follow the same pattern:
the number of limit cycles does not increase
when the nonlinearity $\epsilon$ increases. Moreover,
the bifurcation curves for intermediate ($0\ll\epsilon\ll\infty$)
nonlinearity are always located between the bifurcation curves
corresponding to the two extreme regimes. This means
that although the variation of the nonlinearity $\epsilon$
introduces an important
modification of the time scale and wave form
of the oscillation, it perturbs slightly
its amplitude. Only if two limit cycles have
a very close amplitude for a given $\epsilon$,
there exists the possibility of
collapse of those limit cycles by a saddle-node
bifurcation when $|\epsilon|$ increases.
If the system loses these two limit cycles
it do not recover them for a stronger nonlinearity $\epsilon$.
If the amplitudes of the limit cycles are separated
enough for a given $\epsilon$, the number
of periodic motions is conserved
when $\epsilon$ is varied. \newline
In particular, if we restrict ourselves
to even-polynomial viscous forces,
this behaviour suggests that Lins-Melo-Pugh
conjecture on the number of limit
cycles of Li\'{e}nard systems is true.
This is so because the conjecture is true
in the weakly nonlinear regime and, according
to the behaviour above explained, it should
be true for any other regime.

{\bf Acknowledgements:}  We aknowledge Prof. J. Sesma for
useful discussions and continous encouragement.
We thank the CICYT (Spanish Governement) for finantial support.

\newpage
\begin{center} {\bf Figure Captions} \end{center}

{\bf Figure 1}: A typical phase portrait of Eq. (3).
The limit cycles of amplitudes $a_i$, $i=1,2,\cdots$,
enclose the origin and have the symmetry
$(x,y)\leftrightarrow (-x,-y)$.
Stable and unstable limit cycles alternate.

{\bf Figure 2}: Qualitative bifurcation diagram in the parameter plane
$(k,\epsilon)$ of system (2) for $f(x)$ given in Example 2.
In region $I$, where $k<0$, there is only one periodic solution;
in region $II$, where $0<k<k_*(\epsilon)$ or $k>k^*(\epsilon)$,
there are two limit cycles, and in region $\bigcirc$, where
$k_*(\epsilon)<k<k^*(\epsilon)$, there are none. On the line
$k=0$ the system undergoes an Andronov-Hopf bifurcation
and on the curves $k_*(\epsilon)$ and $k^*(\epsilon)$ a
saddle-node bifurcation. (Nomenclature in the text:
$k_*(0)\equiv k_0$, $k^*(0)\equiv k^0$,
$k_*(\infty)\equiv k_{\infty}$ and $k^*(\infty)\equiv k^{\infty}$).

{\bf Figure 3}: The complete bifurcation diagram of system (2)
for $f(x)$ given in Example 3. The system has no
periodic solutions in region $\bigcirc$, one limit cycle in
region $I$ and two limit cycles in region $II$.
On the line $\delta=0$ the system undergoes an Andronov-Hopf
bifurcation, and on the curves
$B_0(\delta,\mu)=9\mu^2-40\delta=0$
and $B_{\infty}(\delta,\mu)=\mu^2-5\delta=0$
a saddle-node bifurcation
arises for $\epsilon=0$ and $\epsilon=\infty$, respectively.
In region $B$ are located all the bifurcation curves
$B_{\epsilon}(\delta,\mu)$: the curve $B_{\epsilon_2}$
is located between $B_{\epsilon_1}$ and $B_{\infty}$
if $\epsilon_1<\epsilon_2$.

{\bf Figure 4}: Bifurcation curves of system (2) for
$f(x)$ given in Example 4:
{\bf (a)} Amplitude bifurcation-diagram where $a_i^0(b)$
correspond to $\epsilon=0$ and $a_i^{\infty}(b)$
to $\epsilon=\infty$, $i=1,2,3$. The amplitude-bifurcation
$a_i^{\epsilon}(b)$ curves are located in the interior
of the shaded region for every $\epsilon$. The number and
amplitudes of limit cycles for $b=\bar{b}$ are the number
and $a$-coordinates of the intersections between the
curves $a_i^{\epsilon}(b)$ and the line $b=\bar{b}$.
(solid lines correspond to stable cycles and dashed lines
to unstable ones for $\epsilon>0$). \newline
{\bf (b)} Qualitative bifurcation diagram
in the parameter plane
$(b,\epsilon)$. In region $I$ there is only a
periodic solution and in region $III$ there
are three limit cycles.

\end{document}